\documentclass[
reprint,
superscriptaddress,
amsmath,
amssymb,
aps,
pra,
]{revtex4-1}
\usepackage{graphicx}
\usepackage{dcolumn}
\usepackage{bm}
\usepackage[update,prepend]{epstopdf}
\usepackage[colorlinks,
			linkcolor=blue,
            anchorcolor=blue,
            citecolor=blue,
            urlcolor=blue,
            breaklinks=true]{hyperref}

\begin{document}

\title{Inhomogeneity-related cutoff dependence of the Casimir energy and stress}

\author{F. Bao}
\affiliation{Department of Physics, Zhejiang University, Hangzhou 310058, China}
\affiliation{Centre for Optical and Electromagnetic Research, JORCEP, Zhejiang University, Hangzhou 310058, China}
\author{J. S. Evans}
\affiliation{Centre for Optical and Electromagnetic Research, JORCEP, Zhejiang University, Hangzhou 310058, China}
\author{M. Fang}
\affiliation{Centre for Optical and Electromagnetic Research, ZJU-SCNU Joint Research Center of Photonics, South China Normal University, Guangzhou 510000, China}
\author{S. He}\email{sailing@kth.se}
\affiliation{Centre for Optical and Electromagnetic Research, JORCEP, Zhejiang University, Hangzhou 310058, China}
\affiliation{Centre for Optical and Electromagnetic Research, ZJU-SCNU Joint Research Center of Photonics, South China Normal University, Guangzhou 510000, China}
\affiliation{Department of Electromagnetic Engineering, Royal Institute of Technology, 10044 Stockholm, Sweden}
\date{\today}

\begin{abstract}
The cutoff dependence of the Casimir energy and stress is studied using the Green's function method for a system that is piecewise-smoothly inhomogeneous along one dimension. The asymptotic cylinder kernel expansions of the energy and stress are obtained, with some extra cutoff terms that are induced by the inhomogeneity. Introducing interfaces to the system one by one shows how those cutoff terms emerge and illuminates their physical interpretations. Based on that, we propose a subtraction scheme to address the problem of the remaining cutoff dependence in the Casimir stress in an inhomogeneous medium, and show that the nontouching Casimir force between two separated bodies is cutoff independent. The cancellation of the electric and magnetic contributions to the surface divergence near a perfectly conducting wall is found to be incomplete in the case of inhomogeneity.
\end{abstract}

\maketitle

\section{Introduction\label{sec:intro}}
Decades ago Casimir theoretically predicted the attraction between two neutral, perfectly conducting plates by virtue of the zero-point energy \cite{Casimir1948a}. His seminal work started the exploration of a new territory \cite{Plunien1986,Milton2001,Bordag2009} where the vacuum fluctuations result in previously unexpected effects. The Casimir force was confirmed by later experiments \cite{Mohideen1998,Bressi2002,Lamoreaux2005,Klimchitskaya2009} and might find applications in nano/micro-electromechanical-systems (NEMS/MEMS) \cite{Chan2001a,Emig2007a,Rodriguez2011a,Rodriguez2011}. Despite the wide interest in the Casimir effect, due to its fascinating physical effects such as the quantum repulsion \cite{Boyer1968a,Leonhardt2007,Munday2009} and levitation \cite{Rodriguez2008,Sah2010,Rodriguez2010,Rodriguez2013}, quantum friction \cite{Pendry1997,Kardar1999}, quantum torque \cite{Munday2005,Philbin2008}, and also due to the deep relationship between vacuum energy and the cosmology constant \cite{Weinberg1989}, there remains open questions regarding the cutoff dependence or the divergence in the Casimir energy and stress.

The cutoff terms which need to be renormalized trace back to Casimir's original work where he considers the total vacuum energy and subtracts a term that is already present even without the plates. This term is not subject to the boundary conditions and turns out to be diverging. A similar treatment can be found in the general theory \cite{Dzyaloshinskii1961} of van der Waals force (also Casimir force) by Dzyaloshinskii \textit{et al.}, where they have subtracted from the stress a term that one would get if the medium is unbounded, uniform and of the same local property $ \varepsilon $. This subtraction is based on the assumption that the short waves do not sense the ``inhomogeneity" (changing of $ \varepsilon $ across different objects) and thus do not contribute to the Casimir force. Casimir and Dzyaloshinskii's treatments have successfully predicted a force that agrees quite well with experiments, and thus are widely adopted as an essential procedure (subtraction of the ``empty space", the ``unbound medium", or the ``bare/bulk contribution"). However, this is not always adequate. When the space is abnormal \cite{Leonhardt2007}, or curved geometries are present \cite{Deutsch1979,Bordag1999,Bordag2001a}, the energy or stress will still be diverging after the essential procedure of subtraction, even with the dispersion considered \cite{Bordag2002}. The remaining cutoff dependences make it ambiguous to calculate some physical quantities (the value of the cutoff needs to be determined by experiments) like the surface tension associated with the surface divergence \cite{Candelas1982241}, and some of the cutoff terms even do not seem renormalizable \cite{Bordag1999,Graham2002,Graham2003,Fulling2003}. Fortunately, the Casimir force between separated bodies is not influenced and is always finite (free of cutoff dependence), see the discussions in \cite{Graham2004} and in Sec.~4.3.3 in Ref.~\cite{Bordag2009}. This has led to many successful methods \cite{Milton2008,Sah2009,Emig2007,Kenneth2008,Golestani2009,Rodriguez2007} which concern merely the force between separated bodies enclosed by arbitrary manifolds.

Recently, the remaining divergence in the stress tensor has been reported in the case where two flat half spaces are separated by a continuously inhomogeneous medium \cite{Philbin2010,Simpson2013}, even with dispersion considered. Curiosities mainly lie in the question whether the Casimir force between separated bodies is cutoff dependent or not and the problem on how to obtain a finite and meaningful stress to describe the Casimir interaction. Nevertheless, allowing for inhomogeneity of the medium not only complicates the nontouching Casimir interaction, but also permits the nonvanishing local pressure from the medium on a single compact body, and moreover, implies the non-uniformity of the surface tension at the interfaces. The electromagnetic contribution to the local pressure and the surface tension is cutoff dependent, which is reasonable as it actually relates to the Casimir self interaction and the interaction between two touching objects---the compact body and the surrounding medium. The usual law of the Casimir force $ a^{-4} $ implies this cutoff dependence when the separation $ a\to0 $. In a previous paper \cite{Bao2015}, by considering an infinitesimally thin film so as to neglect the cutoff-dependent pressure 
\footnote{The electromagnetic contribution to the surface tension is independent of the position of the film in the global calculation of the Casimir energy based on the mode-summation method, but it's position-dependent in the local calculation using the Green's function method (the presence of the $ x^{-2} $ term in $ \mathcal{U}_s $). This conflict (appears as the presence or absence of $ t^{-1} $ term in the regularised Casimir energy) can be found in many papers in the literature and has been reported and analysed in Fulling's paper \cite{Fulling2003} in 2003. For more details and references we refer the readers to that paper, and here we just remind that it affects the definition of the surface tension but has nothing to do with the nontouching Casimir interaction.}
, we have shown that the nontouching Casimir force (NCF) is cutoff independent, in the first-order perturbation theory. In this paper, we ignore the dispersion of the media, adopt the Green's function method to study the cylinder kernel expansions of the energy and stress, and propose a modified Dzyaloshinskii's subtraction \cite{Dzyaloshinskii1961} to retrieve the nontouching Casimir interaction (note that the NCF is not an independent observable unless the nonuniform pressure balances out and the surface energy is conserved). The cutoff dependences are analysed term by term and the unresolved renormalization of the quadratic and logarithmic divergences found in curved geometries are discussed here in the context of inhomogeneity.

This paper is organised as follows. In Sec.~\ref{sec:method}, we show in detail how we can get the asymptotic expansion of the Casimir energy, for different setups. In Sec.~\ref{sec:analysis}, we discuss how the cutoff terms change when the setup is modified so that we can assign the physical meanings reasonably. A modified Dzyaloshinskii's subtraction is also given. In Sec.~\ref{sec:application}, we study the cutoff dependence of the Casimir energy and stress for two systems of different kinds of optical response. The cancellation between the electric and magnetic contributions to surface divergence near a perfectly conducting (PEC) wall is discussed. We also show numerically that the stress is finite based on our modified subtraction scheme. And we conclude in Sec.~\ref{sec:conclusion}. Throughout this paper, we use the natural units $ \hbar=c=1 $.

\section{Methods\label{sec:method}}
We study the fluctuating electromagnetic field in a medium that is inhomogeneous along one dimension, say the $ x $ axis. The stress tensor and the energy density can be rewritten in terms of the Green's tensor. Due to the translation invariance along the $ y $-$ z $ plane, the Green's tensor can be reduced to two scalar Green's functions which account for the electric and magnetic contributions, respectively \cite{Philbin2010} \begin{equation}\label{eq:emgreenf}
\left\lbrace
\begin{split}
[\mu\varepsilon\kappa^2+u^2-\partial_x\mu^{-1}\partial_x\mu](\mu^{-1}\tilde{\cal G}_E)=&\delta(x-x'), \\ [\mu\varepsilon\kappa^2+u^2-\partial_x\varepsilon^{-1}\partial_x\varepsilon](\varepsilon^{-1}\tilde{\cal G}_M)=&\delta(x-x'),
\end{split}
\right.
\end{equation}
where $ \mu(x) $ is the permeability, $ \varepsilon(x) $ is the permittivity, $ u $ is the wave number along the $ y $-$ z $ plane, $ \partial_x $ is the full derivative with respect to $ x $, and $ \kappa $ is the imaginary frequency $ \omega=i\kappa $. In this section, we give a general mathematical description of these two scalar Green's functions, and seek the asymptotic expansions of the Casimir energy density and stress. For simplicity, we start with a scalar version of the problem, corresponding to the magnetic contribution when $ \mu=1 $, to highlight the physical importance of the expansion terms, and then give the full results of the electromagnetic problem in Sec.~\ref{sec:application}.

The scalar field equation is $ \left[\partial^2_t - \nabla\cdot\varepsilon^{-1}\nabla \right]\varphi=0 $, and its Hamiltonian is \begin{equation}\label{eq:hamiltonian}
\mathcal{H}=\frac{1}{2}\partial_t\varphi\partial_t\varphi+\frac{1}{2}\varepsilon^{-1}\nabla\varphi\cdot\nabla\varphi.
\end{equation}
For a scalar field interacting with a background by the Lagrangian $ \mathcal{L}_{int}=-\frac{1}{2}\lambda\sigma\varphi^2 $, we refer the readers to Ref.~\cite{Graham2004}. With point-splitting regularization, the vacuum or Casimir energy density is \begin{equation}\label{eq:energyfeynman}
\begin{split}
\mathcal{U}
&= \langle 0| \mathcal{H} |0 \rangle \\
&= \lim\limits_{t\to t' \atop \mathbf{r}\to \mathbf{r'}}{\frac{1}{2}\left[\partial_t\partial'_t + \varepsilon^{-1}\nabla\cdot\nabla'\right] \langle 0| T \varphi(t,\mathbf{r}) \varphi(t',\mathbf{r'}) |0 \rangle} \\
&= \lim\limits_{t\to t' \atop \mathbf{r}\to \mathbf{r'}}{\frac{1}{2i}\left[\partial_t\partial'_t + \varepsilon^{-1}\nabla\cdot\nabla'\right]\Delta(t,\mathbf{r};t',\mathbf{r'})},
\end{split}
\end{equation}
where $ T $ is the time-ordering symbol and $ \Delta $ is the Feynman propagator that satisfies $ \left[\partial_t^2 - \nabla\cdot\varepsilon^{-1}\nabla\right]\Delta=\delta(t-t')\delta(\mathbf{r}-\mathbf{r'}) $. For stationary boundaries, we can use the time-translation invariance to obtain the Fourier image of $ \Delta $, $ \tilde{G}(\mathbf{r},\mathbf{r'},\omega) $, satisfying $ \left[-\omega^2-\nabla\cdot\varepsilon^{-1}\nabla\right]\tilde{G}=\delta(\mathbf{r}-\mathbf{r'}) $. Performing Wick's rotation $ \omega=i\kappa $, finally we have \begin{equation}\label{eq:energygreen}
\mathcal{U}=\frac{1}{2\pi}\int\limits_0^\infty \left[-\kappa^2+\varepsilon^{-1}\nabla\cdot\nabla'\right]\tilde{G} \mathrm{d}\kappa.
\end{equation}
Hereafter, the limit $ \mathbf{r'}\to\mathbf{r} $ is implied in the expression of the Casimir energy density. For a dielectric function that varies along the $ x $ axis, using the space-translation invariance along $ y $-$ z $ plane, the energy density could be simplified further as \begin{equation}\label{eq:energycartesian}
\mathcal{U}=
\frac{1}{4\pi^2} 
\iint \limits_0^{\infty} u\varepsilon^{-1}\left[-\varepsilon\kappa^2+u^2+\partial_x\partial'_x\right] \tilde{\cal G} \mathrm{d}u\mathrm{d}\kappa,
\end{equation}
where $ \tilde{\cal G}(x,x',u,i\kappa) $ meets exactly the bottom definition in Eq.~(\ref{eq:emgreenf}) with $ \mu=1 $.

Consider the differential equation \begin{equation}
\left[ k^2-\partial_x\varepsilon^{-1}\partial_x\varepsilon \right]Y=0 ,
\end{equation}
where $ k^2\equiv u^2+\varepsilon\kappa^2, \varepsilon>0 $ and $ u\equiv k\cos\theta $. When $ \varepsilon $ is smooth and $ \xi^{-1}\equiv \sqrt{u^2+\kappa^2} \gg 1 $ thus the wave number $ k \gg 1 $ (hereafter, $ k\gg 1 $ always means $ \xi^{-1}\gg 1 $), $ Y $ can be approximated by the WKB method to arbitrary order, $ Y=\exp\left\lbrace \frac{1}{\xi}\sum_{n=0}^{\infty}\xi^n S_n(u,\kappa,x) \right\rbrace $. Substituting into the equation yields two types of $ Y $, one of which is $ A=f_1(u,\kappa,x)e^{f_2(u,\kappa,x)} $ and the other is $ B=f_1(u,\kappa,x)e^{-f_2(u,\kappa,x)} $, where $ f_1=\exp\left\lbrace\frac{1}{\xi}\sum_{i=1,3,5,\dots}^{\infty}\xi^i S_i\right\rbrace $ and $ f_2=\frac{1}{\xi}\sum_{j=0,2,4,\dots}^{\infty}\xi^j S_j $. Specifically, we have
\begin{equation}\label{eq:wkb}
\begin{split}
&\dot{S}_0 =\xi k,\\
&\dot{S}_1 =-\frac{\dot{\varepsilon}}{4\varepsilon}(2+\sin^2\theta),\\
&\dot{S}_2\dot{S}_0 = \frac{\dot{\varepsilon}^2}{32\varepsilon^2}(12-5\sin^4\theta)-\frac{(2-\sin^2\theta)}{2^3}\frac{\ddot{\varepsilon}}{\varepsilon},\\
&\begin{split}
\dot{S}_3\dot{S}_0^2 =
&\frac{24+12\sin^2\theta-15\sin^6\theta}{2^{6}}\frac{\dot{\varepsilon}^3}{\varepsilon^3}\\
&-\frac{16+4\sin^2\theta-9\sin^4\theta}{2^{5}}\frac{\dot{\varepsilon}\ddot{\varepsilon}}{\varepsilon^2}\\
&+\frac{2-\sin^2\theta}{2^{4}}\frac{\varepsilon^{(3)}}{\varepsilon},
\end{split}\\
&\begin{split}
\dot{S}_4\dot{S}_0^3 = &\frac{1008+960\sin^2\theta+600\sin^4\theta-1105\sin^8\theta}{2^{11}}\frac{\dot{\varepsilon}^4}{\varepsilon^4}\\
&-\frac{248+196\sin^2\theta+50\sin^4\theta-221\sin^6\theta}{2^{8}}\frac{\dot{\varepsilon}^2\ddot{\varepsilon}}{\varepsilon^3}\\
&+\frac{10+5\sin^2\theta-7\sin^4\theta}{2^{5}}\frac{\dot{\varepsilon}\varepsilon^{(3)}}{\varepsilon^2}\\
&+\frac{28+12\sin^2\theta-19\sin^4\theta}{2^{7}}\frac{\ddot{\varepsilon}^2}{\varepsilon^2}\\
&-\frac{2-\sin^2\theta}{2^{5}}\frac{\varepsilon^{(4)}}{\varepsilon},
\end{split}
\end{split}
\end{equation}
which will be used later. Here dot means the full derivative with respect to $ x $. The exact form of these two solutions is usually unavailable and thus is less interesting to us. Instead, we aim to construct the Green's function $ \tilde{\cal G} $ using them. The result is \begin{equation}\label{eq:greenf}
\left\lbrace
\begin{split}
\tilde{\cal G} &= \tilde{\cal G}_{b1}+\tilde{\cal G}_{b2}+\tilde{\cal G}_s,\\
\tilde{\cal G}_{b1} &=  \frac{1}{\varepsilon' W'}\left[\cal AB'-A'B\right]\left(\Theta(x-x')-\frac 12\right),\\
\tilde{\cal G}_{b2} &=- \frac{1}{2\varepsilon' W'}\left[\cal AB'+A'B\right],\\
\tilde{\cal G}_s &= \frac{1}{\varepsilon' W'} \left(  c_1 \mathcal{A} + c_2 \mathcal{B} \right) .
\end{split}
\right.
%\vspace{1pt}
\end{equation}
Here $ W\equiv A\dot{B}-\dot{A}B=-2f_1^2\dot{f}_2 $ is the Wronskian that has the property $ \partial_x\left(\varepsilon W\right)=0 $, $ \Theta $ is the Heaviside step function, $ c_1 $ and $ c_2 $ are arbitrary constants which are determined by the boundary conditions, $ \mathcal{A}\equiv \varepsilon A $, $ \mathcal{A'}\equiv \varepsilon' A' $, and so on [in the following contents, $ \dot{\cal A}\equiv \partial_x(\varepsilon A) $]. $ \tilde{\cal G}_s $ is sometimes referred to as the scattering Green's function, and $ \tilde{\cal G}_b=\tilde{\cal G}_{b1}+\tilde{\cal G}_{b2} $ referred to as the bare Green's function. Although $ \tilde{\cal G}_{b1} $ alone is a special solution to the point-source equation that $ \tilde{\cal G} $ obeys, it is $ \tilde{\cal G}_b $ that is the well-defined singular part of $ \tilde{\cal G} $, because $ \tilde{\cal G}_{b1} $ and $ \tilde{\cal G}_{b2} $ both diverges when $ k\to\infty $ or $ x-x'\to\infty $ but $ \tilde{\cal G}_{b} $ does not. When the medium is homogeneous, say vacuum, we have $ A=\frac{1}{\sqrt{k}}e^{kx}, B=\frac{1}{\sqrt{k}}e^{-kx} $, and $ W=-2 $, then we have $ \tilde{\cal G}_b=\frac{1}{2k}e^{-k\left|x-x'\right|} $ which is exactly the vacuum Green's function. When the medium is smoothly inhomogeneous, $ \tilde{\cal G}_b $ differs from the bare Green's function for the homogeneous medium of the same local $ \varepsilon $ \cite{Guerin2007}, and consequently the energy density also differs.

\subsection{Smooth inhomogeneity\label{sec:si}}
If the inhomogeneity we consider is smooth and the system is unbounded, $ c_1 $ and $ c_2 $ are both zero. When substituted into Eq.~(\ref{eq:energycartesian}), $ \tilde{\cal G}_{b1} $ always gives a contribution of $ \delta(x-x') $ to the kernel of the integral. The energy density reads
\begin{equation}\label{eq:scalarenergydiffexpr}
\begin{split}
\mathcal{U}_b= &\frac{1}{4\pi^2 \sqrt{\varepsilon}} \int\limits_{0}^{\infty}k^2 \mathrm{d}k\int\limits_{0}^{\pi/2}\mathrm{d}\theta\cos\theta\Biggl\lbrace 
\delta(x-x')\\
&-\left[k^2(\cos^2\theta-\sin^2\theta)+\frac{\dot{\mathcal{A}}\dot{\mathcal{B}}}{\mathcal{AB}}\right]\varepsilon^{-1}\left(\frac{\cal A B}{\varepsilon W}\right) \Biggr\rbrace.
\end{split}
\end{equation}
As can be seen from the WKB approximation, when $ k\gg 1 $, we can expand $ \cal AB $ as $ \frac{\cal AB}{\varepsilon W}=-\frac{\varepsilon}{2}\frac{1}{\dot{f}_2}=\sum_{i=1}^{i=\infty} a_i(x,\theta) k^{-i} $.
Specifically, we have $ a_i=0 $ for all $ i\in Evens$, $ a_1=-\frac{1}{2}\varepsilon $, $ a_3=\frac{\varepsilon}{2}\dot{S}_0\dot{S}_2 $, and $ a_5=\frac{\varepsilon}{2}\left(\dot{S}_0^3\dot{S}_4-\dot{S}_0^2\dot{S}_2^2\right) $. Also we have
\begin{equation}\label{eq:abab}
\begin{split}
\frac{\cal \dot{A} \dot{B}}{\cal A B} =&
-k^2 +\left(\mathcal{S}^2-2\dot{S}_0\dot{S}_2\right)\\
&-\frac{(\dot{S}_0\dot{S}_2)^2-2\dot{S}_0^2\dot{S}_3\mathcal{S}+2\dot{S}_0^3\dot{S}_4}{k^2}+O(k^{-4}),
\end{split}
\end{equation}
where $ \mathcal{S} \equiv \dot{S}_1+\partial_x\ln\varepsilon $. Introducing an upper cutoff of the wave number $ \Lambda=\delta(x-x') $ to the integral (note the lower bound is not zero because we require $ k\gg 1 $), or alternatively using the cutoff parameter $ t=1/\Lambda=\min\{x-x'\} $, and leaving aside the cumbersome calculations, we have the regularized Casimir energy density
\begin{widetext}
\begin{equation}\label{eq:scalarenergyrslt}
\begin{split}
\mathcal{U}^{reg}_b=
&\frac{1}{16\pi^2 \sqrt{\varepsilon}} t^{-4}+\frac{1}{4\pi^2\sqrt{\varepsilon}}\frac{504\varepsilon\ddot{\varepsilon}-509\dot{\varepsilon}^2}{6720\varepsilon^2} t^{-2}\\
&-\frac{1}{4\pi^2\sqrt{\varepsilon}}\frac{1152\varepsilon^3\varepsilon^{(4)}-5280\varepsilon^2\dot{\varepsilon}\varepsilon^{(3)}-4240\varepsilon^2\ddot{\varepsilon}^2+15680\varepsilon\dot{\varepsilon}^2\ddot{\varepsilon}-7155\dot{\varepsilon}^4}{30720\varepsilon^4} \ln t +O(1).
\end{split}
\end{equation}
\end{widetext}
The leading term is the energy density for a homogeneous medium, which can be easily verified by the mode-summation method \begin{equation}\label{eq:energyhomogeneous}
\mathcal{U}_h^{reg}=\frac{1}{2}\sum_J{\omega_J}=\frac{1}{2\sqrt{\varepsilon}}\int_{0}^{1/t}{\frac{\mathbf{d} \mathbf{k}^3}{(2\pi)^3}k}=\frac{1}{16\pi^2 \sqrt{\varepsilon}} t^{-4}.
\end{equation}
The subleading terms are nonzero when the medium is inhomogeneous.

\subsection{One interface\label{sec:so}}
Consider the inhomogeneity which is smooth everywhere except at $ x=0 $. Denote $ L $ the left domain $ x<0 $, $ I $ the interface, and $ R $ the right domain $ x>0 $. When $ x,x'>0 $, along with the direct propagation from $ x' $ to $ x $, now the scattering takes place $ \tilde{\cal G}^R=\tilde{\cal G}_b+\tilde{\cal G}_s^R $. Using the continuity of $ \tilde{\cal G} $ and $ \varepsilon^{-1}\partial_x \tilde{\cal G} $ at the interface, we can obtain from Eq.~(\ref{eq:greenf}) the scattering Green's function
\begin{equation}\label{eq:scalarscattering}
\tilde{\cal G}_s^R = \frac{\mathcal{B}^R \mathcal{B}'^R}{\varepsilon^R W^R}\frac{\mathcal{A}^R_0 }{\mathcal{B}^R_0}r^{RL}_0,
\end{equation}
where
\begin{equation}\label{eq:reflectioncoefficient}
r^{RL}_0\equiv \left[\frac{\varepsilon^R\partial_x\ln\mathcal{A}^L-\varepsilon^L\partial_x\ln\mathcal{A}^R}{\varepsilon^R\partial_x\ln\mathcal{A}^L-\varepsilon^L\partial_x\ln\mathcal{B}^R}\right]_0.
\end{equation}
Hereafter subscript 0 (or $ a $ in the following contents) means the quantity is evaluated in the limit $ x\to 0 $ (or $ a $) from the right ($ RL $) or left ($ LR $) side. In the other domain where $ x,x'<0 $, the scattering Green's function is just like the above with all $ L $ and $ R $ interchanged, and all $ \cal A $ and $ \cal B $ interchanged. For clarity, the superscripts in the following formulas will be suppressed if they can be figured out by the context.

The quantity $ r $ is actually the generalised reflection coefficient at the interface. Expanded when $ k\gg 1 $, $ r^{RL}_0 $ gives a leading term that is identical to the \textit{p}-polarised Fresnel formula \begin{equation}
r^{(0)}=-\left[\frac{\varepsilon^Lk^R-\varepsilon^Rk^L}{\varepsilon^Rk^L+\varepsilon^Lk^R}\right]_0.
\end{equation}
Its first subleading term is \begin{equation}\label{eq:reflection1}
r^{(1)}=-\frac{2\varepsilon^L_0k^R_0 (\varepsilon^L\mathcal{S}^R-\varepsilon^R\mathcal{S}^L )_0}{(\varepsilon^Rk^L+\varepsilon^Lk^R)^2_0}.
\vspace{1pt}
\end{equation}
The second and third subleading terms are given in Appendix~\ref{app:reflctn}. All the higher terms are unnecessary for analysing the cutoff dependence.

Substituting the scattering Green's function into Eq.~(\ref{eq:energycartesian}) and letting $ x\to x' $, we get
\begin{widetext}
\begin{equation}\label{eq:scalarsurfaceenergydensity}
\mathcal{U}_s^R=
\frac{1}{4\pi^2} \iint\limits_{0}^{\infty}u
\left\lbrace 
(u^2-\varepsilon\kappa^2)\frac{\mathcal{B}}{\mathcal{A}} + \frac{\cal \dot{B} \dot{B}}{\cal A B} 
\right\rbrace \frac{\mathcal{A}_0}{\mathcal{B}_0}r_0 \frac{\mathcal{A}\mathcal{B}}{\varepsilon W} \varepsilon^{-1} \mathrm{d}u\mathrm{d}\kappa,
\end{equation}
where
\begin{equation}\label{eq:bbab}
\frac{\cal \dot{B} \dot{B}}{\cal A B}\frac{\mathcal{A}_0}{\mathcal{B}_0} = \left[
k^2 -2k\mathcal{S}+\mathcal{S}^2+2\dot{S}_0\dot{S}_2-\frac{2}{k}(\dot{S}_0\dot{S}_2\mathcal{S}+\dot{S}_0^2\dot{S}_3)+O(k^{-2})
\right]e^{-2f_2|_0^x},
\end{equation}
\end{widetext}
and $ \frac{\mathcal{B}}{\mathcal{A}}\frac{\mathcal{A}_0}{\mathcal{B}_0}=e^{-2f_2|_0^x} $. The exponent can be expanded as
\begin{equation}\label{eq:f2expansion}
-2f_2|^x_0 = -2\xi^{-1}S_0|^x_0 -2\xi S_2|^x_0-2\xi^3 S_4|^x_0 +O(\xi^5).
\end{equation}
The leading term $ -2\xi^{-1}S_0|^x_0=-2\int_{0}^{x}k(\bar{x})\mathrm{d}\bar{x} $ will be kept in the exponent while the subleading terms can be pulled down by an exponential expansion.
% \begin{equation}
%\begin{split}
%&\exp\lbrace-2\xi S_2|^x_0-2\xi^3 S_4|^x_0\rbrace\\
%=& 1-2\xi S_2|^x_0+(2\xi S_2|^x_0)^2-(2\xi S_2|^x_0)^3-2\xi^3 S_4|^x_0 +O(\xi^4).
%\end{split}
%\end{equation}
Recall the expansion of $ \frac{\cal AB}{\varepsilon W} $ and $ r_0 $, we see there are four different types of $ k $ that have been used in the energy density---$ k_0,k(\bar{x}),k(x) $ and $ \xi^{-1} $. To unify the notation, we define
\begin{equation}
\left\lbrace
\begin{aligned}
k(\bar{x})&=\sqrt{\cos^2\theta_0+\frac{\varepsilon(\bar{x})}{\varepsilon_0}\sin^2\theta_0}\times k_0 &\equiv \mathcal{N}(\bar{x},0)k_0,\\
k(x)&=\sqrt{\cos^2\theta_0+\frac{\varepsilon(x)}{\varepsilon_0}\sin^2\theta_0}\times k_0 &\equiv \mathcal{N}(x,0)k_0,\\
\xi^{-1} &=\sqrt{\cos^2\theta_0+\frac{1}{\varepsilon_0}\sin^2\theta_0}\times k_0 &\equiv \mathcal{N}(\infty,0)k_0,
\end{aligned}
\right.
\end{equation}
where $ \cal N $ indicates the transformation of the wave number (an analogue of the refractive index). Now the energy density becomes
\begin{equation}\label{eq:scalarenergydensitykintegral}
\mathcal{U}_s\sim \sum\limits_{n=1}^{5}\int_{0}^{\pi/2}\int_{k_c}^{\infty}k_0^{4-n}e^{-2k_0\mathcal{X}}\mathrm{d}k_0\mathcal{F}(\theta_0,x)\mathrm{d}\theta_0+O(1),
\end{equation}
where $ \cal F $ is some function of $ \theta_0 $ and $ x $, and $ \mathcal{X}\equiv \int_{0}^{x} \mathcal{N}(\bar{x},0)\mathrm{d}\bar{x} $ is an analogue of the optical distance. When $ x\to 0 $ so $ \mathcal{X}\to 0$, the $ k $ integral in Eq.~(\ref{eq:scalarenergydensitykintegral}) yields five diverging terms $ \mathcal{X}^{-4},\mathcal{X}^{-3},\mathcal{X}^{-2},\mathcal{X}^{-1} $ and $ \ln\mathcal{X} $ (precisely these five terms come with a factor $ e^{-2k_c\mathcal{X}} $, but this factor doesn't affect the diverging property; the last term $ \ln\mathcal{X} $ is integrable near the interface). The divergence of the energy density when the interface is approached, the so-called surface divergence, is a well-known phenomenon \cite{Deutsch1979,KENNEDY1980346,Milton2006}. Specifically, the term $ \mathcal{X}^{-4} $ relates to $ r^{(0)} $ while the term $ \mathcal{X}^{-1} $ involves up to $ r^{(3)} $ and thus $ \varepsilon^{(3)} $. Therefore, the surface divergence would disappear if the continuity of $ \varepsilon^{(0)\sim (3)} $ at the interface is met. A similar result has been reported by K. Milton in Ref.~\cite{Milton2011a}. There, the scalar field in a monomial potential $ [\partial_t^2-\nabla^2+x^\alpha]G=\delta $ is studied. And the surface divergence is found to disappear when $ \alpha>2 $ (though the two results are slightly different because of the different models. They consistently show that, the surface divergence will disappear if the medium is smooth enough).

It is still challenging to get an explicit full expansion of the energy density for an arbitrary permittivity. The difficulty lies in getting function $ \cal F $. To simplify the problem while still focusing on the influence of the inhomogeneity, we keep only the first two terms in the above exponential expansion, so that $ \exp\lbrace-2f_2|^x_0\rbrace=e^{-2k_0\mathcal{X}}(1-2\xi S_2|^x_0) $. By doing this, we are able to get the first subleading term of the expansion of the energy density. Besides, we let $ \varepsilon^L\to\infty $ so that the reflection coefficient is also simplified
\begin{equation}\label{eq:neumann}
r^{RL}_0=-1-2\mathcal{S}_0/k_0.
\end{equation}
Furthermore, after the $ k $ integral in Eq.~(\ref{eq:scalarenergydensitykintegral}), we take the near-interface limit $ x\ll 1 $ so that $ \mathcal{N}(x,0)=1+\frac{1}{2}(\sin^2\theta\partial_x\ln\varepsilon)_0x,\mathcal{X}=x[1+\frac{1}{4}(\sin^2\theta\partial_x\ln\varepsilon)_0x] $ and $ S_2|^x_0=(\dot{S}_2)_0x $. At last, the energy density is
\begin{equation}\label{eq:scalarsurfaceenergydensityr}
\mathcal{U}_s^{R}=\frac{1}{16\pi^2\sqrt{\varepsilon_0}}\left\lbrace x^{-4}-\frac{7}{60}\left(\frac{\dot{\varepsilon}}{\varepsilon}\right)_0x^{-3}\right\rbrace+O(x^{-2}).
\vspace{1pt}
\end{equation}

If it is $ \varepsilon^R\to\infty $, we have $ \mathcal{A} \leftrightarrow \mathcal{B} $ and $ L\leftrightarrow R $ those interchanges in Eq.~(\ref{eq:scalarsurfaceenergydensity}). But in an elegant way we will have the same result as Eq.~(\ref{eq:scalarsurfaceenergydensityr}).

\subsection{Two interfaces\label{sec:st}}
When the medium contains two interfaces at $ x=0,a $, letting $ L,C $ and $ R $ denote the left, central and right domains respectively, the scattering Green's function in the central domain takes the form
\begin{widetext}
\begin{equation}\label{eq:scalarscattering2}
\tilde{\cal G}_s^{LR} = \frac{1}{\varepsilon W}\frac{1}{1-\mathcal{P}r^{CL}_0r^{CR}_a}\left\lbrace
\left[\mathcal{AA'}\left(\frac{\mathcal{B}}{\mathcal{A}}\right)_ar^{CR}_a+\mathcal{BB'}\left(\frac{\mathcal{A}}{\mathcal{B}}\right)_0r^{CL}_0\right]-(\mathcal{AB'+A'B})\mathcal{P}r^{CL}_0r^{CR}_a
\right\rbrace,
\end{equation}
\end{widetext}
where $ \mathcal{P}\equiv\left(\frac{\mathcal{A}}{\mathcal{B}}\right)_0\left(\frac{\mathcal{B}}{\mathcal{A}}\right)_a=e^{-2f_2|^a_0} $ describes the one-round propagation of the photon in the central domain.

Physically, the photon propagator from source $ x' $ to point $ x $ where $ 0<x,x'<a $, includes the contributions of direct propagation and reflections by the interfaces. The former gives the bare Green's function, while the latter gives the scattering Green's function. However, the single reflection by one interface accounts for the interaction between two adjoining domains, and does not contribute to the nontouching interaction between two separated objects which intuitively should involve multiple reflections between two interfaces. Therefore, from Eq.~(\ref{eq:scalarscattering2}) we should subtract Eq.~(\ref{eq:scalarscattering}) and a counterpart for the interface at $ a $, \textit{i.e.}, $ \tilde{\cal G}_c=\tilde{\cal G}_s^{LR}-\tilde{\cal G}_s^L-\tilde{\cal G}_s^R $. Then we get the Green's function that describes the nontouching Casimir interaction we are interested in
\begin{widetext}
\begin{equation}\label{eq:scalarcasimirgreenf}
\tilde{\cal G}_c = \frac{1}{\varepsilon W}\frac{1}{1-\mathcal{P}r^{CL}_0r^{CR}_a}\left\lbrace
\left[\mathcal{AA'}\left(\frac{\mathcal{B}}{\mathcal{A}}\right)_ar^{CR}_a+\mathcal{BB'}\left(\frac{\mathcal{A}}{\mathcal{B}}\right)_0r^{CL}_0\right]-(\mathcal{AB'+A'B})\right\rbrace\mathcal{P}r^{CL}_0r^{CR}_a.
\end{equation}
\end{widetext}
The factor $ \cal P $ is essential to make the energy density $ \mathcal{U}_c $ (yielded by $ \tilde{\cal G}_c $) finite near the interface. To see this, we don't need to do the asymptotic expansion, but note that it is proportional to $ \exp\left[-2\int_{0}^{a}k \mathrm{d}x\right] $, an exponentially decaying term that makes any polynomial integrable over $ k $. It follows that, the nontouching Casimir force is cutoff independent.

\section{Analyses\label{sec:analysis}}
By using the macroscopic quantity $ \varepsilon $ to describe the medium and its interaction with the field, we actually imply the possibility of self interaction of the medium mediated by the field. A well-known picture of that is, the medium can be polarized by a photon by an operator $ \hat{\Pi}=\frac{\varepsilon-1}{4\pi}\omega^2\delta(\mathrm{r}-\mathrm{r'}) $, and the free energy is thought of as the interacting energy between polaritons mediated by virtual photons \cite{Dzyaloshinskii1961}. Not only the dielectrics, even the vacuum can produce electron-positron pairs and is thus polarizable. In light of these, the Casimir energy given by Eq.~(\ref{eq:energyfeynman}) or (\ref{eq:emenergydensity}) can be understood as a part of self energy of the medium. Specifically, according to the previous section, when the medium is smooth and unbounded, the energy density is $ \mathcal{U}_b $. Thus it corresponds to the self interaction of an unbounded smooth medium. When there is one interface, the energy density is $ \mathcal{U}^{R(L)}=\mathcal{U}_b+\mathcal{U}_s^{R(L)} $. The newly arising energy $ \mathcal{U}_s^{R(L)} $ thus corresponds to the interaction of two adjoining domains. When there are two interfaces, the energy density is
\begin{equation}\label{eq:energysplit}
\mathcal{U}^{LR}=\mathcal{U}_b+\mathcal{U}_s^L+\mathcal{U}_s^R+\mathcal{U}_c.
\end{equation}
The energy $ \mathcal{U}_c $ corresponds to the nontouching interaction between two separated domains.

Let us focus on the case of two interfaces. The first three terms of Eq.~(\ref{eq:energysplit}) are all cutoff dependent. To remove the infinities, Dzyaloshinskii \textit{et al.} proposed to subtract a term that we would get if the medium is unbounded and homogeneous with the same local property $ \varepsilon $, that is, the leading term in Eq.~(\ref{eq:scalarenergyrslt}), based on the assumption that short waves do not sense the inhomogeneity and thus do not contribute to the Casimir force. This is reasonable for real materials, and at least works for homogeneous media. In the latter case, all the subleading terms in Eq.~(\ref{eq:scalarenergyrslt}) vanish, and $ \mathcal{U}_s^{R(L)} $ is independent of the position thus vanishes in calculating the force, leaving us only $ \mathcal{U}_c $ that we are actually interested in [Note that if we are considering the EM field in a homogeneous medium, $ \mathcal{U}_s^L+\mathcal{U}_s^R $ (and similarly stress) will vanish near a PEC boundary due to the cancellation between the electric and magnetic contributions, as shown in Eq.~(\ref{eq:emsurface1})]. Again, in the case of the homogeneous medium, the Dzyaloshinskii's subtraction has removed all the energy that is not subject to the boundary conditions (the bare Green's function contribution), equivalent with Casimir's original treatment.

When the medium is smoothly inhomogeneous and non-dispersive, arbitrarily short waves contribute to the Casimir force, so the Dzyaloshinskii's subtraction fails. In order to fix this, Philbin \textit{et al.} proposed removing the contributions of the small-scale inhomogeneity by using the Green's function given by the first-order WKB approximation as Eqs.~(36) and (37) in Ref.~\cite{Philbin2010} (up to $ S_0 $ and $ S_1 $). Here we see, this cannot remove completely all cutoff terms caused by $ \mathcal{G}_b $ for an arbitrary $ \varepsilon $, but indeed relaxes the divergences. To remove $ \mathcal{U}_b $ from $ \cal U $, we need higher-order approximations up to $ S_4 $.

The last term of Eq.~(\ref{eq:energysplit}) is called the free Casimir energy, because it can be released by changing the separation distance, giving rise to the nontouching Casimir force. However, when the medium is inhomogeneous, we see the energy $ \mathcal{U}_s^{R(L)} $ will also change along with the shift of the interface, invalidating the usual definition of the free Casimir energy as the energy difference when the separation is finite and when the separation is infinite $ E_c=E(a)-E(\infty) $ (invalidating Casimir's treatment). Consequently, the nontouching Casimir force is no longer an independent observable unless the integral of the scattering energy over all interfaces, here the sum $ \mathcal{U}_s^L+\mathcal{U}_s^R $, gives a constant. Notably, for the EM field, the electric and the magnetic contributions now do not cancel completely for PEC boundaries, in contrast to the case of a homogeneous medium. To retrieve the free Casimir energy (stress), we propose the following subtraction scheme \begin{equation}\label{eq:subtraction}
\mathcal{U}_c=\mathcal{U}^{LR}+\mathcal{U}_b-\mathcal{U}^L-\mathcal{U}^R,
\end{equation}
or one can directly use the corresponding Green's function as Eq.~(\ref{eq:scalarcasimirgreenf}).

When the energy density is integrated over the whole space, we get the well-known cylinder kernel expansion of the Casimir energy \cite{Fulling2003}. Homogeneous (sectionally) medium and flat boundaries would yield the simplest result $ E\sim Vt^{-4}+St^{-3}+E_c+O(t) $ where the inverse cubic term can still be zero when some kind of cancellation takes place. The presence of curvature of the boundaries \cite{Deutsch1979,Bordag2001a,Bordag2002} would immediately introduce some extra terms so that $ E\sim Vt^{-4}+St^{-3}+e_2t^{-2}+e_3t^{-1}+f\ln t+E_c+O(t) $. Here all the coefficients are related to the geometric characteristics of the system. Particularly, $ V $ is the volume of the space, $ S $ is the surface area of the boundaries, and the others have more complex meanings. According to dimensional analysis, $ e_2 $ should be some length, $ e_3 $ is a pure number, and $ f $ is the inverse of a length. To renormalize these cutoff dependent energies, one could introduce the classical energy of the system $ E^{class}=pV+\gamma S+Fe_2+ke_3+hf $ and absorb $ t $ in these classical quantities $ p,\gamma,F,k$ and $ h $ \cite{Bordag1999,Bordag2009}. However, an unresolved problem is, there seems to be no standard physical meanings for $ F,k $ and $ h $ ($ p $ being the pressure and $ \gamma $ being the surface tension).

Here in our case, the interfaces are flat, and these extra cutoff terms are caused by the inhomogeneity of the medium. Integrating the self energy $ \mathcal{U}_b $ over the space, we won't have an explicit factor $ V $ which we would have if the medium is homogeneous (and we can have if we use some averaged quantity), and also there are quadratic and logarithmic terms. Besides, the integration of the interacting energy $ \mathcal{U}_s $ would yield $ S(t^{-3}+t^{-2}+t^{-1}+\ln t) $ but not only $ St^{-3} $. In this way, we see that, though the volume term $ Vt^{-4} $ must still come from $ \mathcal{U}_b $ and the surface term $ St^{-3} $ must be still from $ \mathcal{U}_s $, the other terms especially the quadratic and logarithmic terms are also from $ \mathcal{U}_b $ and $ \mathcal{U}_s $, and contain contributions from both of them. Note that those terms involve higher-order derivatives of $ \varepsilon $, thus together with $ S $ or under the volume integration, they have coefficients of indeed the dimensions same as $ e_2, e_3 $ or $ f $, respectively.

Based on the above analyses, we want to take a heuristic interpretation of these cutoff terms---terms in $ e_2t^{-2}+e_3t^{-1}+f\ln t $ are not to be renormalized by some new classical quantities, but are only higher-order corrections to the pressure $ p $ and the surface tension $ \gamma $. That is, introduce $ E^{class}=\int_\Omega p \mathrm{d}V+\int_{\partial\Omega} \gamma \mathrm{d}S $, and absorb the cutoff terms in $ \mathcal{U}_b $ by $ p $ and in $ \mathcal{U}_s $ (precisely $ \int\mathcal{U}_s\mathrm{d}x $) by $ \gamma $. The physics is clear. The self interaction of one object contributes to its inner pressure, and the interaction of two adjoining objects contributes to the surface tension at the interface.

At last, we should mention that, the energy density exactly at the interface is not well-defined, though the Green's function is still continuous there. This is obvious as can be seen in Eq.~(\ref{eq:scalarenergyrslt})---to define the energy density we need the continuity of up to $ \varepsilon^{(4)} $. This might be overcome by softening the interface layer by specifying a smooth profile of the permittivity. It needs future study.

\section{Applications\label{sec:application}}
In this section we apply the above method to the electromagnetic field and give the asymptotic expansions of the energy density and stress, to discuss some interesting topics. Notably, the Minkowski's stress tensor $ \sigma^M_{ij}=D_iE_j+B_iH_j-\frac{1}{2}\delta_{ij}\left( \mathbf{D\cdot E}+\mathbf{B\cdot H} \right) $ and the Raabe-Welsch stress \cite{Raabe2005}, the ordinary energy density of the electromagnetic field $ \mathcal{U}=\frac{1}{2}\int_{0}^{\infty}\left( \mathbf{D\cdot E}+\mathbf{B\cdot H} \right)_\xi \mathrm{d}\xi  $ and the Casimir energy density \cite{Rosa2011} derived from the interaction of dipoles all show similar asymptotic behaviours with minor differences in the coefficients. Therefore, we give only the results of the Minkowski's stress and the energy density derived from it. The component of the stress that generates force is
\begin{widetext}
\begin{equation}\label{eq:stress}
\sigma_{xx}=\frac{1}{4\pi^2\sqrt{\mu\varepsilon}}\int\limits_{0}^{\infty}k^2\mathrm{d}k\int\limits_{0}^{\pi/2}\left\lbrace
\mu^{-1}[k^2-\partial_x\partial'_x]\tilde{\cal G}_E + \varepsilon^{-1}[k^2-\partial_x\partial'_x]\tilde{\cal G}_M
\right\rbrace\cos\theta\mathrm{d}\theta,
\end{equation}
and the energy density is \begin{equation}\label{eq:emenergydensity}
\begin{split}
\mathcal{U}=\frac{1}{4\pi^2\sqrt{\mu\varepsilon}}\int\limits_{0}^{\infty}k^2\mathrm{d}k\int\limits_{0}^{\pi/2}
&\left\lbrace\mu^{-1}[k^2(\cos^2\theta-\sin^2\theta)+\partial_x\partial'_x]\tilde{\cal G}_E \right. \\
&\left. + \varepsilon^{-1}[k^2(\cos^2\theta-\sin^2\theta)+\partial_x\partial'_x]\tilde{\cal G}_M
\right\rbrace\cos\theta\mathrm{d}\theta.
\end{split}
\end{equation}
\end{widetext}
Those two scalar Green's functions are given in Eq.~(\ref{eq:emgreenf}) and have similar solutions as Eq.~(\ref{eq:greenf}) with $ \mu $ or $ \varepsilon $ in appropriate places. Note also that for electric and magnetic Green's functions we have different $ A's $ and $ B's $. In principle, arbitrary permittivity and permeability that vary along one dimension are tractable, but the results are sufficiently cumbersome that only by adding further constrains can we get relatively simple analytical expressions. Two examples we will discuss below are the trivial magnetic response system $ \mu=1 $ and the constant light speed system $ \mu\varepsilon=1 $. In the results, we keep only the first two terms of the expansions.

\subsection{Trivial magnetic response system\label{sec:emmu1}}
When $ \mu=1 $, the WKB approximation for the magnetic Green's function is the same as Eq.~(\ref{eq:wkb}), while for the electric Green's function we have \begin{equation}\label{eq:wkb1te}
\left\lbrace
\begin{split}
&\dot{S}_0 =\xi k,\\
&\dot{S}_1 =-\frac{\dot{\varepsilon}}{4\varepsilon}\sin^2\theta,\\
&\dot{S}_2\dot{S}_0 = \frac{\ddot{\varepsilon}}{8\varepsilon}\sin^2\theta-\frac{5\dot{\varepsilon}^2}{32\varepsilon^2}\sin^4\theta.
\end{split}
\right.
\end{equation}

The bare parts of the regularized energy density and stress are \begin{equation}\label{eq:embare1}
\begin{split}
\mathcal{U}_b^{reg} &=\frac{1}{8\pi^2\sqrt{\varepsilon}}t^{-4}+\frac{1}{4\pi^2\sqrt{\varepsilon}}\frac{224\varepsilon\ddot{\varepsilon}-229\dot{\varepsilon}^2}{3360\varepsilon^2}t^{-2}+O(\ln t),\\
\sigma_b^{reg} 
&=-\frac{1}{24\pi^2 \sqrt{\varepsilon}} t^{-4}-\frac{1}{4\pi^2\sqrt{\varepsilon}}\frac{23\dot{\varepsilon}^2}{480\varepsilon^2}t^{-2}+O(\ln t).
\end{split}
\end{equation}

When there is one interface at $ x=0 $ and $ \varepsilon_L\to\infty $ in this or the constant light speed system, the electric Green's function obeys the Dirichlet boundary condition while the magnetic Green's function obeys the Neumann condition. The latter condition yields the same reflection coefficient as Eq.~(\ref{eq:neumann}), while the Dirichlet one yields exactly $ r_E=1 $. The fact that $ r_E $ has an opposite sign against $ r_M $, to some extent, results in the cancellation between the electric and magnetic contributions. The scattering parts of energy and stress read \begin{equation}\label{eq:emsurface1}
\begin{split}
\mathcal{U}_s^R
&=\frac{-1}{16\pi^2\sqrt{\varepsilon_0}} \left(\frac{7\dot{\varepsilon}}{60\varepsilon}\right)_0x^{-3}+O(x^{-2}),\\
\sigma_s^R
&=\frac{1}{16\pi^2\sqrt{\varepsilon_0}}
\left(\frac{5\dot{\varepsilon}}{12\varepsilon}\right)_0x^{-3}
+O(x^{-2}).
\end{split}
\end{equation}
The absence of the $ x^{-4} $ term in the energy density is due to the above mentioned EM cancellation, while we see when the medium is inhomogeneous, the cancellation is incomplete. The absence of that term in the stress, however, is a generic property, because the operator $ \partial_x\partial_x' $ acting on the scattering Green's function would produce a leading term $ k^2 $ so that to the leading term, $ (k^2-\partial_x\partial_x') $ vanishes in the expression of the stress. Again, we see when the medium is inhomogeneous, there are subleading terms. And, this is exactly the reason to the divergence near the interfaces reported in \cite{Philbin2010}. As analysed in the previous section, here we give a numeric illustration of the proposed subtraction scheme [using Eqs.~(\ref{eq:scalarcasimirgreenf}) and (\ref{eq:stress})] to fix the problem raised by the above reference.
\begin{figure}[t]
\scalebox{0.85}{\includegraphics{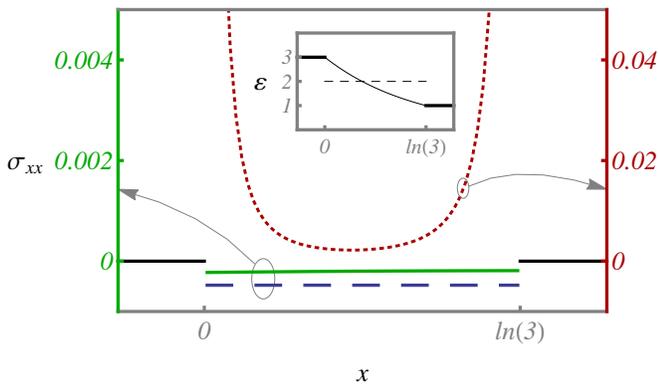}}
\caption{\label{fig:stress}The dotted red curve recovers the previous result \cite{Philbin2010}. While using our proposed subtraction scheme, we obtain the solid green curve. The dashed blue curve is the stress when $ \varepsilon=2 $. The inset is the profile of $ \varepsilon $.}
\end{figure}

As can be seen in Fig.~\ref{fig:stress}, the profile of the permittivity is plotted in the inset [$ \varepsilon^L=3,\varepsilon^R=1 $, and $ \varepsilon^C=3e^{-x} $ (solid) or $ \varepsilon^C=2 $ (dashed)]. The dotted red curve in the main figure is simulated with $ \mathcal{\tilde{G}}^{LR}-\mathcal{\tilde{G}}_b $ (the previous method, where $ \mathcal{B} $ and $ \cal A $ are the first and second kind of the modified Bessel functions, respectively) which still contains $ \mathcal{\tilde{G}}_s^L+\mathcal{\tilde{G}}_s^R $ so that the stress diverges near the interfaces. Using $ \mathcal{\tilde{G}}_c $, immediately we arrive at the solid green curve. The dashed blue curve is for the case $ \varepsilon^C=2 $. Note that, either of these two $ \varepsilon^C $ satisfies the Casimir repulsion criterion $ \varepsilon^L>\varepsilon^C>\varepsilon^R $, thus it's reasonable to see the solid green curve in the same side (negative side of the vertical axis) with the dashed blue curve (while the dotted red curve is in the positive side that corresponds to attraction).

\subsection{Constant light speed system\label{sec:emmuep1}}
When $ \mu\varepsilon=1 $, $ \xi k=1 $. Thus for the magnetic Green's function, we have \begin{equation}\label{eq:wkbim}
\left\lbrace
\begin{split}
&\dot{S}_0 =1,\\
&\dot{S}_1 =-\frac{\dot{\varepsilon}}{2\varepsilon},\\
&\dot{S}_2\dot{S}_0 =\frac{3\dot{\varepsilon}^2}{8\varepsilon^2}-\frac{\ddot{\varepsilon}}{4\varepsilon}.
\end{split}
\right.
\end{equation}
Making a substitution $ \varepsilon\to\varepsilon^{-1} $, we can get the approximation of the electric Green's function expressed by $ \varepsilon $ \begin{equation}\label{eq:wkbie}
\left\lbrace
\begin{split}
&\dot{S}_0 =1,\\
&\dot{S}_1 =\frac{\dot{\varepsilon}}{2\varepsilon},\\
&\dot{S}_2\dot{S}_0 =-\frac{\dot{\varepsilon}^2}{8\varepsilon^2}+\frac{\ddot{\varepsilon}}{4\varepsilon}.
\end{split}
\right.
\end{equation}

The bare parts of the regularized energy density and stress are \begin{equation}\label{eq:embarei}
\begin{split}
\mathcal{U}_b^{reg}
&=\frac{1}{8\pi^2}t^{-4}+\frac{1}{4\pi^2}\frac{\dot{\varepsilon}^2}{24\varepsilon^2}t^{-2}+O(\ln t),\\
\sigma_b^{reg}
&=-\frac{1}{24\pi^2 } t^{-4}-\frac{1}{4\pi^2}\frac{\dot{\varepsilon}^2}{8\varepsilon^2}t^{-2}+ O(\ln t).
\end{split}
\end{equation}

The scattering parts are \begin{equation}\label{eq:emscatteringi}
\begin{split}
\mathcal{U}_s^R
&=\frac{-1}{16\pi^2}\left(\frac{\dot{\varepsilon}}{3\varepsilon}\right)_0x^{-3}+O(x^{-2}),\\
\sigma_s^R
&=\frac{1}{16\pi^2}\left(\frac{\dot{\varepsilon}}{\varepsilon}\right)_0x^{-3}+O(x^{-2}).
\end{split}
\end{equation}

This constant light speed system is interesting because, the usual surface divergence near a curved boundary was reported to disappear \cite{Brevik1982179,Brevik1983237}. Here, we see the inhomogeneity-induced surface divergences are still present in this constant light speed system.

\section{Conclusions\label{sec:conclusion}}
To summarise, we have studied the asymptotic cylinder kernel expansion of the Casimir energy and stress within an smoothly inhomogeneous medium, using the WKB approximation of the Green's function. When the medium consists of several parts and thus is piecewise-smoothly inhomogeneous, we have interpreted the self energy of the medium as possessed by the interaction among those parts. A heuristic renormalization of the cutoff terms has been given. Also, we have given a modified subtraction scheme to retrieve a finite stress to describe the nontouching Casimir interaction which consequently is finite as well, with numeric illustration. Moreover, we have studied the cancellation between the electric and magnetic contributions to the surface divergence near a PEC wall, and the surface divergence for the constant light speed system. For the former, we have found the cancellation is incomplete when the medium is inhomogeneous. And for the latter, we have found the surface divergence is still present, in contrast to the case of curvature.

\appendix
\begin{widetext}
\section{Reflection coefficients\label{app:reflctn}}
Following Eq.~(\ref{eq:reflection1}), the second order correction of the reflection coefficient is
\begin{equation}
r^{(2)}=-\frac{2\varepsilon^L_0k^R_0 (\varepsilon^L\mathcal{S}^R-\varepsilon^R\mathcal{S}^L )^2_0}{(\varepsilon^Rk^L+\varepsilon^Lk^R)^3_0}-2\varepsilon^L_0\varepsilon^R_0\frac{(\dot{S}^L_0\dot{S}^R_2-\dot{S}^R_0\dot{S}^L_2)_0}{(\varepsilon^Rk^L+\varepsilon^Lk^R)^2_0},
\end{equation}
and the third order reads
\begin{equation}
\begin{split}
r^{(3)}=&-\frac{2\varepsilon^L_0k^R_0 (\varepsilon^L\mathcal{S}^R-\varepsilon^R\mathcal{S}^L )^3_0}{(\varepsilon^Rk^L+\varepsilon^Lk^R)^4_0}-\frac{2\varepsilon^L_0(\varepsilon^L\mathcal{S}^R-\varepsilon^R\mathcal{S}^L )_0\left[\varepsilon^R(\dot{S}^L_0\dot{S}^R_2-\dot{S}^R_0\dot{S}^L_2)-\dot{S}^R_0(\varepsilon^L\dot{S}^R_2+\varepsilon^R\dot{S}^L_2)\right]_0}{(\varepsilon^Rk^L+\varepsilon^Lk^R)^3_0}\\
&-\frac{2\varepsilon^L_0k^R_0 (\varepsilon^L\dot{S}_3^R\dot{S}_0^L\dot{S}_0^R-\varepsilon^R\dot{S}_3^L\dot{S}_0^L\dot{S}_0^R)_0}{k^L_0k^R_0(\varepsilon^Rk^L+\varepsilon^Lk^R)^2_0}.
\end{split}
\end{equation}
\end{widetext}

\begin{acknowledgments}
This work is partially supported by the Science and Technology Department of Zhejiang Province and the Guangdong Innovative Research Team Program (No.201001D0104799318).
\end{acknowledgments}

%\nocite{*}
\bibliography{Casimir}

\end{document}